\documentclass[3p,preprint,authoryear]{elsarticle}

\usepackage{amssymb}
\usepackage{graphicx}
\biboptions{comma}

\usepackage[figuresright]{rotating}

\begin{document}
\pagestyle{empty}

\begin{frontmatter}


%% \title{Convenient liquidity measure for financial markets\tnoteref{label1}}

 \title{An instantaneous market volatility estimation}

\author{Oleh Danyliv}
\author{Bruce Bland}
\address{Fidessa group plc, One Old Jewry, London, EC2R 8DN,
United Kingdom} \tnotetext[t1]{Authors are grateful to Dr. Daniel
Nicholass, Dr. Christian Voigt, Jon Davidson at Fidessa group plc
for valuable discussions and support.}
\tnotetext[t2] {The views expressed in this article are those of the authors and do not necessarily reflect the views
of Fidessa group plc, or any of its subsidiaries.}

\begin{abstract}
Working on different aspects of algorithmic trading we empirically
discovered a new market invariant. It links together the
volatility of the instrument with its traded volume, the average
spread and the volume in the order book. The invariant has been
tested on different markets and different asset classes. In all
cases we did not find significant violation of the invariant. The
formula for the invariant was used for the volatility estimation,
which we called the {\it instantaneous volatility}. Quantitative
comparison showed that it reproduces realised volatility better
than one-day-ahead GARCH(1,1) prediction. Because of the
short-term prediction nature, the instantaneous volatility could
be used by algo developers, volatility traders and other market
professionals.\end{abstract}

\begin{keyword}
volatility \sep order book \sep realised volatility \sep market
invariant

\JEL G12 \sep G14 \sep G17

\end{keyword}

\end{frontmatter}

\pagestyle{headings} \setcounter{page}{1}
%\pagenumbering{roman}

%% Introduction
\section{Introduction}
\label{introcution}

Predicting and understanding of financial market volatility is
central to the theory and practice of asset pricing, risk
management, optimal order execution. Standard calculation of
historical volatility uses log price returns over some time
horizon. Different models of ARCH family  could be applied to this
data to make some volatility forecast. Unfortunately, these
historical estimations and forecasts are biased and very sensitive
to data outliers. An infamous example of such a bias is given by
\citep{figlewski1994}: the market crash in October 19, 1987 caused
a huge increase in estimated volatility to around 27 percent,
although the implied volatility quickly dropped to a usual level
of 15 percent a few days later. This left market participants with
the dilemma to use either a new "historical" estimate or to be
more consistent with the option pricing pre-crash value.

From another point of view, an algo trading requires a short term
volatility estimation, which changes significantly throughout the
day. For example, the volatility of UK stocks could increase by
50\% at the start of the trading session in the US. That cannot be
predicted by a calculation of a daily historical volatility and
requires building an intraday volatility profile, similar to the
volume profile used in benchmark VWAP algos. Working in this area
and trying to improve the performance of algos, we discovered a
new way of volatility estimation. It comes from the fact, that the
price move and the trading activity affect the order book in a
predictable way. Using this property, we derived the formula for
instantaneous volatility which requires only a short term market
observation. It  is not based on a specific model, nor on the
historical calculation, but solely relies on a new market
invariant, which links together volatility, traded volume, order
book volume and the spread. We will explain the way we discovered
the invariant and will show that the invariant holds for liquid
markets. At the end of the paper we will compare our data to
realised volatility \citep{andresen2001} and GARCH(1,1) forecast.

Our analysis is based on a one tick quote and trade data for
liquid stocks of European indexes  London Stock Exchange (FTSE All
Shares and FTSE 100), Tokyo Stock Exchange (Nikkei 225), Frankfurt
Stock Exchange (DAX 30), Nasdaq Stockholm (OMX 30) and Toronto
Stock Exchange (S\&P/TSX 60) in 2016. Derivatives data on S\&P
E-mini, oil contracts, US Treasuries and German bonds correspond
to the end of 2016/ start of 2017 period. The data was provided by
Fidessa's High Performance Trade Database of the Analytical
Framework and in-house High Performance Quote Database.

%% \section{Small order execution time}

\section{Small order execution time}
\label{section:SmallOrderEX}

An execution time of an order, which is placed on a touch level
(top bid price for buy orders and ask price for sell orders) is an
important practical problem which arises in broker and algo
trading. Since the market price might "run away" from the order
level, this problem does not have a solution all the time and a
more accurate formulation of the problem would be "given a maximum
order  waiting time $t$ what is the average trading time of a
passively executed limit order with a fixed limit price?" This
problem is quite complex for real stocks and derivatives, but
there are two extreme cases when it is possible to advance with
estimations. First of all, it is the case of a limit for a
volatile instrument whose price action can be described by a
random walk: in this case a queue of the order could be neglected.
We will call this case the Execution by Price. The second extreme
is the case of an order for a low volatile instrument. In this
case the only way for an order to get executed is through waiting
its turn in the order queue (we call it the Execution by Trading
Activity).

\subsection{Execution by Price}
Let us consider the price action of an instrument which could be
described as a random walk: the price of this instrument moves up
and down with equal probability. If $\sigma(\Delta T)$ is the standard
deviation of the random walk during the measurement period $\Delta T$ ,
then for an arbitrary time $t$, the volatility follows the square
root scaling rule

\begin {equation}
    \sigma(t) = \sigma(\Delta T) \sqrt{\frac{t}{\Delta T}}.
    \label{scaling}
\end {equation}

In order to have a good chance of a passive execution, the
obtained value of the standard deviation should be of the same
level of magnitude as the spread: the price needs cross the spread
in order to fill the order passively and if the spread is too wide
(comparing to the volatility during the waiting time $t$), passive
order executions will be rare. On the other side, if the waiting
time is too big and $\sigma(t)$ is much larger than the historical
average of the spread $\left< spread\right>$, then a passive
execution becomes very probable, but the risk grows. It is a risk
of a very bad execution when, trying to capture a small spread,
trader loses much larger value $\sigma(t)$: the opportunity cost
of the execution becomes very high. Therefore, the time at which
the standard deviation of the price is equal to the average spread
is an important characteristics of any passive order execution. It
is logical to denote this time as $T_{Price}$ since it is depends
purely on the price action:
\begin {equation}
    T_{Price}= \Delta T \left( \frac{\left<spread\right>}{\sigma(\Delta T)} \right)^2
    \label{t_price}
\end {equation}

It should
be noted that $T_{Price}$ is not equal to an average waiting time of
an executed limit order. It could be shown analytically \citep{danyliv2015}
that for the binary random walk, waiting
this amount of time would
correspond to the probability $p = 1-{\rm erf} (\frac{1}{\sqrt{2}})$ or 32\% of a passive
execution of the order, placed on the touch level.

\subsection{Execution by Trading Activity}

If the volatility of the instrument is low, the limit order can
still be filled if the trading activity is high. If during a
sample time $\Delta T$ the amount of the traded volume was
$V_{Traded}(\Delta T)$, then, in the equilibrium condition, half
of these trades will happen on bid and half of them will take
place on ask levels and the volume traded on one side of the
market during time $t$ is:

\begin {equation}
    V(t)=\frac{t}{2} \times \frac{V_{Traded}(\Delta T)}{\Delta T}.
    \label{v_t}
\end {equation}

To have a plausible chance of a passive execution, this traded
volume should be comparable to the length of the order queue. For
a buy order, placed on the best bid price level, the average queue
size is the average volume on the bid level $\left<
V_{BID}\right>$. The estimation of the queue size which is
independent of the trade direction is the average of bid and ask
volumes $\frac{\left< V_{BID}\right> + \left< V_{ASK}\right>
}{2}$. Therefore, the characteristic time $T_{Volume}$ in which a
limit order will be traded on the market could be defined as

\begin {equation}
    T_{Volume} =
    \Delta T \left(\frac{\left< V_{BID}\right> + \left< V_{ASK}\right>}{V_{Traded}(\Delta T)}\right)
    \label{t_volume0}
\end {equation}

Unfortunately, in reality the situation is slightly more complex
because for instruments with a wide spread, trades could happen
not just on the best bid/offer level, but also inside the spread.
Therefore, not all traded volume should be taken into account, but
only $V_{Traded}(\Delta T)\times P$ part of it. For buy orders,
the correction coefficient $P$ is the probability of trades to
take place on the order level before the order is filled. Because
the price could move in small increments called tick size ($TS$),
the ratio $n \equiv \frac{\left<spread\right>}{TS}$ will
correspond to the number of price levels the price can jump to.
The more the number of such states, the more likely that a trade
will happen there and less probable that the trade will eliminate
the queue in front of the limit order. That is why the correction
coefficient is likely to be a function of the spread size in ticks
between bid and ask levels. If the spread is minimal ($n=1$),
there is no chance for trades to be executed inside the spread,
$P(1)=1$ and formula (4) does not need correction. For very large
spreads we can assume that the trades are normally distributed
around bid/ask prices and only half of all trades will eliminate
the queue, setting $P(\infty)=\frac{1}{2}$.

\begin{figure}[htp]
    \centering
    \includegraphics[width=0.7\textwidth]{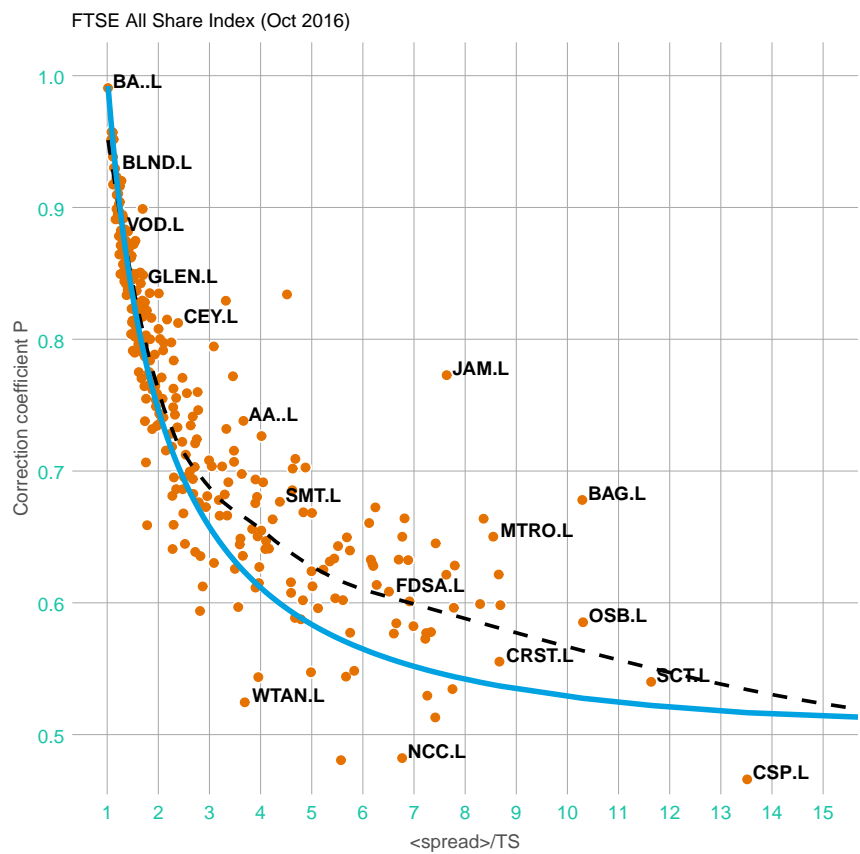}
        \caption{The probability of trades to participate in an order
        queue depletion  during time (\ref{t_volume0}) as a function of
        the average spread.
        Orange dots show real data for London All Share stocks,
        blue line is the prediction given by formula (\ref{correction_coeff}).
        Thin dashed line corresponds to the local polynomial regression fitting. }
    \label{figure:correction}
\end{figure}

The most consistent way of checking how much traded volume is
participating in the queue depletion process is to make direct
simulations of limit orders and then count how much volume is
traded at the level of initial touch price and below (for buy
orders). The results of such simulations for stocks of London
Stock Exchange (LSE) are shown on Fig.\ref{figure:correction}.
Each dot on this chart is one month's worth of limit order
simulations for one instrument. To obtain an analytic formula for
the correction coefficient, one can assume that the probability of
a trade declines exponentially with the distance to the initial
touch level. The analysis of data showed that the power of the
decaying exponent is close to -0.5 or $P \propto
\exp^{-\sqrt{n}}$, where $n$ is the spread size in ticks. Then the
probability of the volume to trade on touch or below, which
satisfies boundary conditions $P(n=1)=1$ and
$P(n=\infty)=\frac{1}{2}$, will have the form

\begin {equation}
    P(n) = \frac{1}{2} \left( 1+ \exp^{- \frac{ n -1}{\sqrt{n}}} \right)
    \label{correction_coeff}
\end {equation}

The predictive power of this formula is shown on
Fig.\ref{figure:correction}, where the correction coefficient
(\ref{correction_coeff}) is represented by the blue solid line. It
works perfectly well for instruments with small spreads and does
not deviate significantly for stocks with large spreads.

Using these results, formula (\ref{t_volume0}) could be corrected
and has the final form

\begin {equation}
    T_{Volume} =
    \Delta T \left(\frac{\left< V_{BID}\right> + \left< V_{ASK}\right>}{V_{Traded}(\Delta T)}\right)
    \frac{2}{1+ \exp^{- \frac{{\left<spread\right>}/{TS} -1}
    {\sqrt{{\left<spread\right>}/{TS}}}}}
    \label{t_volume}
\end {equation}
As in the previous case of the time related to the price, this is
a characteristic time of the execution of low volatile instruments
and does not directly correspond to the average trade time.

%%\section{The Market Invariant}

\section{The Market Invariant}
\label{section:invariant}

\begin{figure}[htp]
    \centering
    \includegraphics[width=0.7\textwidth]{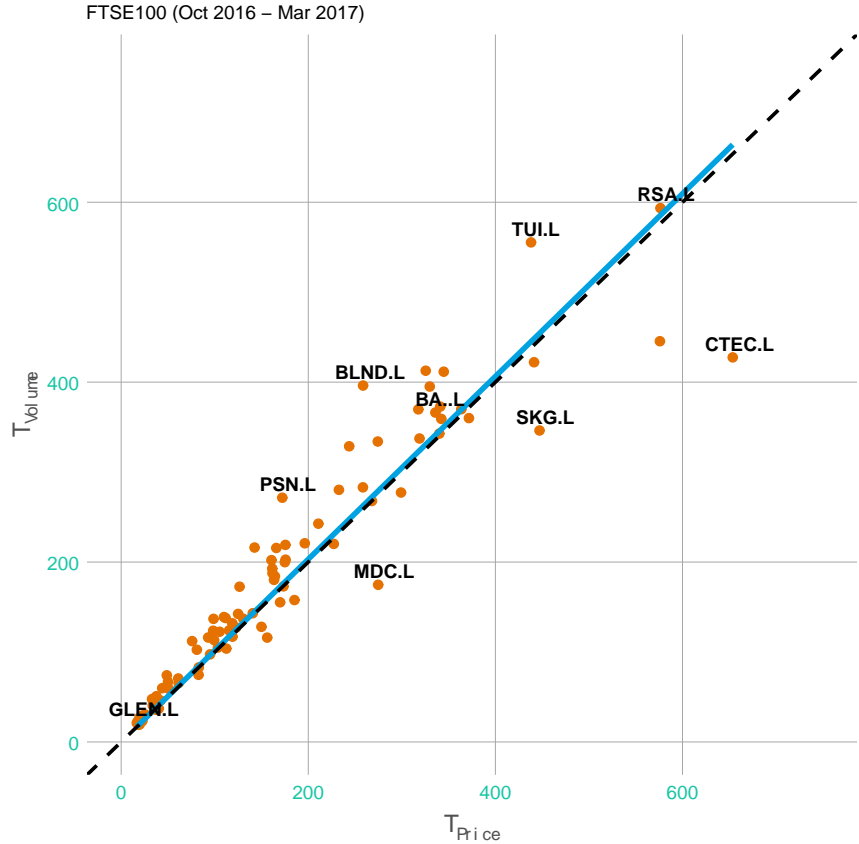}
        \caption{Times $T_{Volume}$ and $T_{Price}$ in seconds for stocks of FTSE100.
        Each orange dot corresponds to the individual instrument.
        Dashed line is a diagonal line, solid blue line corresponds to the regression line from
        forced to cross (0,0) point. }
    \label{figure:lix_sp500_nikkei}
\end{figure}

The characteristic times $T_{Price}$  and $T_{Volume}$ were
derived from different perspectives, but they explain similar
property of the market: they related to a time, which the market
participant has to wait to trade one share. Practical calculations
revealed that they have very similar absolute values.
Fig.\ref{figure:lix_sp500_nikkei} shows averaged over last quarter
of year 2016  time values (2) and (6) for the most liquid stocks
of  London Stock Exchange. The characteristic waiting times range
from around 20 seconds for Glencore (GLEN.L) to 10 mins for RSA
Insurance Group (RSA.L), but both times are very similar with the
correlation coefficient equal to 0.944. Similar analysis for the
same stocks in the first quarter of 2017 gave similarly high
correlation value of 0.902.

Using these observations, one can assume that the
following invariant is present on the market

\begin {equation}
    \gamma^2 \equiv \frac{T_{Volume}}{T_{Price}} = 1.
    \label{invariant0}
\end {equation}

The square root of this ratio will depend linearly from the
standard deviation of the price $\sigma(\Delta T)$, which will be
used further for the volatility estimation. Therefore the
following form of the invariant could have practical
implementation.

\begin {equation}
 \gamma \equiv  \frac{\sigma(\Delta T)}{\left< spread\right>}
 \times \sqrt{\frac{\left< V_{BID}\right> + \left< V_{ASK}\right>}{V_{Traded}(\Delta T)}}
 \times \sqrt{\frac{2}{1+ \exp^{- \frac{{\left<spread\right>}/{TS} -1}
    {\sqrt{{\left<spread\right>}/{TS}}}}}} = 1
    \label{invariant}
\end {equation}

Expression (\ref{invariant}) links together easily measurable
volumes and spread with the standard deviation of the price. In a
nutshell, this expression states the obvious: the volume of the
passive orders and the trading activity will influence the price
fluctuations and the spread of the instrument. Initially the
invariant (8) was tested on highly liquid derivatives such as US
Treasury Notes, S\&P 500 E-minis, WTI crude oil and German
government bonds. Although future contracts could be traded for
years, they are becoming active near the expiration date. To
eliminate illiquid periods, only days with the trading volume
higher than 20\% of the maximum observed traded value for the
contract were taken into consideration to build data sets.

Standard deviation of the price on intraday timescale is tightly
linked to realised volatility $\sigma_R$ \citep{andresen2001},
defined as sum of squared log returns $r_t=\ln (Price(t+ \Delta
T)/Price(t))$. The price of an asset usually does not change
significantly during the day and could be replaced by the daily
average value and an average daily return is roughly zero.
Therefore

\begin {equation}
    r_t=\ln\left( \frac{Price(t+ \Delta T)}{Price(t)}  \right)
    \approx \frac{Price(t + \Delta T) - Price(t)}{Price(t)}
    \approx \frac{Price(t + \Delta T) - Price(t)}{\left <Price\right
    >}
    \label{return}
\end {equation}
and
\begin {equation}
    {\sigma_R} \equiv \sqrt{\sum_t{{r_t}^2}} \approx
    \frac{\sigma(\Delta T)}{\left <Price\right >},
    \label{realised}
\end {equation}
which demonstrates mentioned relation. Five minute intervals were
used for calculations with the overnight return being omitted as
is often done in the literature \citep{brownlees2011}. Since the
realised volatility estimates the daily volatility, the averages
in (\ref{invariant}) should correspond to daily averages. The
results of $\gamma$ calculations are  shown in Table
\ref{table:gamma_futures}, where number of days reflect the size
of the data set.

\begin{table}[ht]
\caption{Testing invariant $\gamma=1$ for derivatives}
% title of Table
\centering % used for centering
\begin{tabular}{| l | c | c | c | c | c | c | c | c |}% centered columns (4
\hline\hline
Instrument & Month &  Days  &  $T_{Price}$ & $\left<
\gamma \right>$ & $\sigma_\gamma$ &  Correlation
    & \multicolumn{2}{c|}{p-value of tests}\\
    \cline{8-9}
    && &sec&&&$\left<
T_{Price},T_{Volumre}\right>$&S-W&K-S\\
\hline

S\&P E-mini        & SEP 2016 & 71 & 26.23 & 0.850 & 0.122 & 0.807 & 0.029 & 0.908 \\
S\&P E-mini        & DEC 2016 & 71 & 20.43 & 0.832 & 0.084 & 0.899 & 0.990 & 0.999 \\
Crude Oil WTI      & DEC 2016 & 42 &  7.27 & 0.905 & 0.132 & 0.461 & 0.001 & 0.123 \\
Crude Oil WTI      & JAN 2017 & 33 &  5.31 & 0.978 & 0.197 & 0.434 & 0.909 & 0.970 \\
US 10-Year T-Note  & JUN 2016 & 84 & 144.5 & 0.885 & 0.110 & 0.777 & 1.2*10$^{-6}$ & 0.060 \\
US 10-Year T-Note  & SEP 2016 & 86 & 136.1 & 0.919 & 0.111 & 0.679 & 0.194 & 0.974 \\
US 10-Year T-Note  & DEC 2016 & 83 & 135.6 & 0.862 & 0.092 & 0.867 & 0.463 & 0.783 \\
German Euro-Bund   & JUN 2017 & 65 & 29.07 & 0.966 & 0.126 & 0.802 & 0.089 & 0.644 \\
German Euro-Buxl   & JUN 2017 & 63 &  32.1 & 1.125 & 0.151 & 0.657 & 0.361 & 0.383 \\

\hline \hline
\end{tabular}
\label{table:gamma_futures}% is used to refer this table in the text
\end{table}

A strong version of null hypothesis ``$\left< \gamma \right > =
1$", which would prove the invariant directly, does not pass the
statistical significance test and should be rejected.
Nevertheless, the validity of the invariant in forms
(\ref{invariant0}) and (\ref{invariant}) could be seen from the
following: first of all, the instrument's mean is very close to
the value one: the distance to this value is never larger than 0.2
(less than 20\%). This is quite a respectable accuracy, taking
into account that one of the variables in the formula for the
invariant is volatility, whose coefficient of determination $R^2$,
according to \cite{koopman2004}, ranges from $0.34$ to $0.6$
depending on predictive model. In terms of standard deviation, six
out of nine observations are within one sigma distance from the
expected value one; for all observations $|1-\left< \gamma
\right>| < 2\sigma_\gamma$. Secondly, we observe strong
correlation between two characteristic times (apart from oil
contracts for which the correlation drops below 0.5).

It is known, that realised volatility itself strongly depends on
the time interval on which it is calculated which might lead to
overestimation or underestimation of real value. If the volatility
in (\ref{invariant}) is overestimated, that will make $\left<
\gamma \right > > 1$ and underestimation will make it smaller than
one. That is why, to eliminate the volatility calculation bias, we
might use a weaker null hypothesis, which states that ``$\gamma$
values are normally distributed". From the results we know that
the expected value is approximately one, but this is not part of
the hypothesis. New null hypothesis was examined by Shapiro-Wilk
(S-W) and Kolmogorov-Smirnov (K-S) tests. The $p$-values of
testing methods should be higher than the cut-off value
$\alpha=0.05$ and would mean that the null hypothesis is not
rejected. For all derivatives gamma values are normally
distributed according to Kolmogorov-Smirnov test. Shapiro-Wilk
tests reject two data sets. The results for crude oil show how
fragile the normality test is: January contract has a good
normality fit although December contract does not pass
Shapiro-Wilk normality test.  From these results we might conclude
that random variable $\gamma $ is likely to be normally
distributed around an expected value close to one.

Additionally, ratio (\ref{invariant}) was tested on a set of
stocks which are part of major indices traded on different venues
around the globe: London Stock Exchange (FTSE All Shares and FTSE
100), Tokyo Stock Exchange (Nikkei 225), Frankfurt Stock Exchange
(DAX 30), Nasdaq Stockholm (OMX 30) and Toronto Stock Exchange
(S\&P/TSX 60). From all the constituencies of an index, only
liquid stocks with $T_{Price}<15$ min  were selected for the
analysis. The quote and trade data for the last quarter of year
2016 was processed and a three month average of $T_{Volume}$ and
$T_{Price}$  where calculated for each stock. It should be noted,
that the resulting data for FTSE 100 was already shown on Fig.
\ref{figure:lix_sp500_nikkei}. Analysing similar charts for other
indexes, we observed that if a stock is under some stressful
condition (earnings, corporate news, reorganisation), the value of
gamma might differ significantly from the expected value of one. A
later chapter will provide evidence for this statement.

For equities we additionally combined the averages for individual
instruments into an exchange average. The results of these
calculations are presented in Table \ref{table:gamma_equities}. An
over line $\overline{\left< \gamma \right>}$ means an additional
exchange average over gamma values over individual instruments.
This value is very close to the value one: $|1-\overline{\left<
\gamma \right>}| < \sigma_{\overline{\gamma}}$ for all indices,
but DAX 30. We also observe, a strong correlation between two
characteristic times for equity indexes which ranges from 0.676
for Canadian stocks to 0.954 to German stocks. Similarly to the
case of derivatives, we also could expect a normal distribution of
$\left< \gamma \right>$ values . According to Kolmogorov-Smirnov
test, the hypothesis of normal distribution is not rejected for
all indices, but Nikkei 225. The Shapiro-Wilk test additionally
disqualifies Swedish OMX 30. It should be noted that the normality
test is very sensitive to outliers: few stocks in a distressed
state could create a bias for the whole exchange. Nikkei stocks,
for example, do not satisfy the normality test, although they show
very close to unity $\left< \gamma \right>$ value and strong
correlation between characteristic times.

\begin{table}[ht]
\caption{Testing invariant $\gamma=1$ for equities}
% title of Table
\centering % used for centering
\begin{tabular}{| l | c | c | c | c | c | c | c | c|}% centered columns (4
\hline\hline Index & \multicolumn{2}{c|}{Instrumens}  &
$T_{Price}$ & $\overline{\left< \gamma \right>}$ &
$\sigma_{\overline{\gamma}}$ & Correlation
    & \multicolumn{2}{c|}{p-value of tests}\\
    \cline{2-3}\cline{8-9}

& Total & $T_{Price} < 15 min$ &sec&&&$\left<
T_{Price},T_{Volumre}\right>$&S-W&K-S\\
\hline

FTSE All Share & 630 & 253  & 338.4 & 0.938 & 0.154 & 0.848 & 0.083& 0.518 \\
FTSE 100       & 100 & 95   & 169.7 & 1.060 & 0.093 & 0.944 & 0.224 & 0.854\\
Nikkei 225     & 225 & 200  & 210.4 & 1.007 & 0.175 & 0.898 & 1.8*10$^{-12}$ &0.012\\
DAX 30         & 30  & 30   & 78.8  & 1.115 & 0.100 & 0.954 & 0.126 & 0.870\\
OMX 30         & 30  & 30   & 210.4 & 1.169 & 0.186 & 0.678 & 7.8*10$^{-6}$ & 0.196\\
S\&P/TSX 60    & 60  & 34   & 95.2  & 1.264 & 0.299 & 0.676 & 0.340 & 0.709\\

\hline \hline
\end{tabular}
\label{table:gamma_equities}% is used to refer this table in the text
\end{table}

Overall we could state that the market invariant (\ref{invariant})
holds for statistical averages in a wide range of markets. The
only condition which we used for the stocks selection process was
a high liquidity of instruments which was expressed as a
relatively low (less than 15 min) characteristic time.

%%\section{Instantaneous Volatility Estimation}
\section{Instantaneous volatility estimation}
\label{section:instatnt}

The volatility estimator $\sigma_I$ during period $\Delta T$ could
be calculated from the standard deviation of the price used in
(\ref{t_price}) via

\begin {equation}
    \sigma_I(\Delta T) \equiv \frac{\sigma(\Delta T)}{\left< Price
    \right>},
    \label{volatility0}
\end {equation}
where angle brackets, as previously, mean historical average.
 Using the market invariant
(\ref{invariant}), the volatility on interval $\Delta T$ could be
estimated as

\begin {equation}
    \sigma_I(\Delta T) = \frac{\left<spread\right>}{\left< Price\right>}
    \sqrt{\frac{V_{Traded}(\Delta T)}{\left< V_{BID}\right> + \left< V_{ASK}\right>}}
    \sqrt{\frac{1}{2} \left(1 + \exp^{- \frac{{\left<spread\right>}/{TS} -1}
    {\sqrt{{\left<spread\right>}/{TS}}}}\right)}.
    \label{volatility}
\end {equation}

Comparing definition (\ref{volatility0}) with the approximation
(\ref{realised}) for realised volatility, it is obvious that
$\sigma_I$ is a proxy for realised volatility. All values on the
right hand side of formula (\ref{volatility}) (apart from traded
volume) do not significantly change with time. Traded volume on
short time intervals could be expressed via trading rate, which
also does not change significantly on a minute to minute basis.
Therefore, a volatility for liquid instruments could be estimated
from a very short-time observation, literally few time intervals.
This feature is valuable in algo trading where such calculations
could be used. Because of the short term nature of the estimation,
we called the obtained value an {\it instantaneous volatility}.

Practical calculation showed that formula (\ref{volatility}) is
robust and could be modified to be truly instantaneous: the
average price could be replaced with the last trading price, the
historical average of the spread and the order book volume could
be replaced by the average over 3-5 level of the market depth. If
the trading volume is estimated from the volume profile, then the
volatility could be calculated from the snapshot of the order book
and a traded volume profile, which is usually available on trading
platforms.

\subsection{Volatility dependence on spread}
The first two terms in (\ref{volatility}) are responsible for the
spread dependency of the volatility. It is quite intuitive, that
the volatility is proportional to the spread: if the price does
not move, trades will take place on static best bids and best
offers which differ by the spread value (so called``bid-ask
bounce"). Therefore, the price change during time interval $\Delta
P \propto spread$. The second term in (\ref{volatility}) makes
this dependency slightly smaller and non-linear. The Taylor
expansion around point $\frac{\left<spread \right>}{TS} = 1$ shows
this explicitly
\[
    \sigma_I  \propto
    \left< spread \right> \left (1-\frac{1}{4}
    \left( \frac{\left<spread \right>}{TS}-1 \right)
    \right).
\]
For large spreads, where $\frac{\left<spread \right>}{TS} \gg 1$,
volatility converges to linear spread dependence
\[
    \sigma_I  \propto
    \frac{\left< spread \right> }{\sqrt{2}}.
\]

\subsection{Volatility dependence on volume}

Strong dependence of realised volatility on trading volume is
known from empirical studies. For example,
\cite{bogousslavsky2019}, reported a high correlation of realised
volatility with an intraday turnover (and negative correlation
with the market depth volume) for NYSE, Amex and NASDAQ stocks.

According to (\ref{volatility}), the volatility estimate depends
on volume as
\[
    \sigma_I \propto
    \sqrt{\frac{V_{Traded}(\Delta T)}{\left< V_{BID}\right> + \left< V_{ASK}\right>}}
\]

The dependence on traded volume is easy to explain: if there is no
trading, the price will be static and that will result in no
volatility. In contrary, a large aggressive buy order will create
high trading activity and potentially will increase the price
(volatility).

From another side, if there is a significant amount of volume in
the order book, it will put brakes on price moves: all this volume
has to be traded for the price to move. An extreme example of this
scenario is a large buy limit order which can completely stop a
downside move of the price. Therefore, it is quite logical that
the volatility is inversely proportional to the volume in the
order book.

Markets created a natural test for the volume dependency: in the
UK, shares, which are listed on London Stock Exchange are also
traded on minor exchanges Chi-X, BATS and Turquoise. These shares
have the same ISIN code and are fully fungible. Because there is
no arbitrage, the spot price on all exchanges are the same for the
same instrument, whereas volumes depend on the popularity of the
exchange and could differ by an order of magnitude. According to
formula (\ref{volatility}), the resulting volatility estimations
using data from different exchanges should be comparable and be in
a line with the historical volatility.

\begin{figure}[htp]
    \centering
    \includegraphics[width=0.7\textwidth]{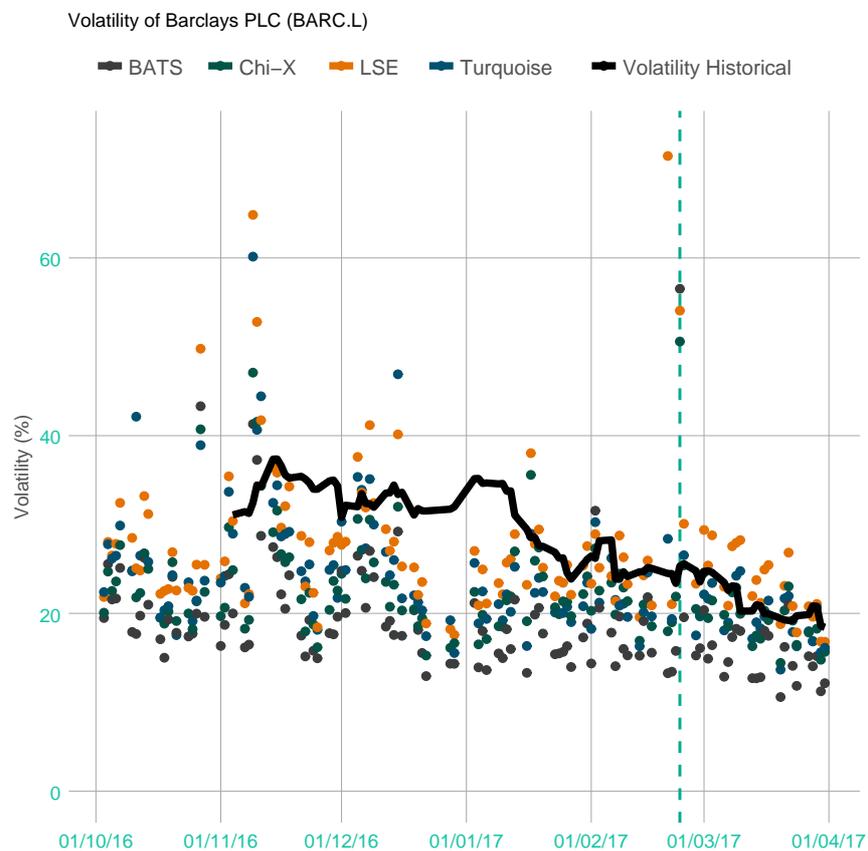}
        \caption{The instantaneous volatility for Barclays PLC,
        calculated on different exchanges is compared to the
        historical volatility (solid line).
        Blue dashed line corresponds to the 23 Feb 2017,
        the day of the annual report.}
    \label{figure:fungible}
\end{figure}

Fig.\ref{figure:fungible} shows that the volatility estimation for
Barclays PLC shares based on the data from different exchanges
produces similar results despite the fact that the volume of
shares traded on LSE for this instrument is 7.8 times higher than
the volume on BATS  and 3 times higher than the volume traded on
Turquoise. This chart also shows strong responsiveness of the
instantaneous volatility: the dashed line on this data shows the
annual report date, when Barclays PLC reported a large increase of
its profit on operations in year 2016.

\subsection{Volatility dependence on time}

The dependence of the instantaneous volatility on the trading
volume implicitly contains a time dependence. Variables like
spread, price and volume in the order book depend on time, but
this time dependence consists only of  some random fluctuations
around fixed constant values. Traded volume, on another hand,
grows linearly with time if the trading rate stays constant:
$V_{Traded}(\Delta T) \propto \Delta T$. This statement is true
when the effect of the volume profile (more active trading at the
beginning and the end of the session) is neglected or the time of
the measurement $\Delta T$ is small. Then the time dependence of
the instantaneous volatility is simply
\[
\sigma_I \propto \sqrt{\Delta T},
\]
which is the expected time scaling for the price volatility
measure, predicted by the random walk model.

%\section{Comparison to realised volatility and GARCH(1,1)}
\section{Comparison to realised volatility and GARCH(1,1)}
\label{section:comparison}

As previously noted, formula (\ref{volatility}) allows an
immediate volatility estimation. It requires a short-time trading
history and order book information, which makes it useful for the
volatility estimations on intraday timeframes. It is difficult to
compare it to historical estimations of volatility since they work
on larger, usually daily or weekly data. As in the case of the
invariant, realised volatility (\ref{realised}) could be used to
quantify the accuracy of the estimation.

\begin{figure}[htp]
    \centering
    \includegraphics[width=0.7\textwidth]{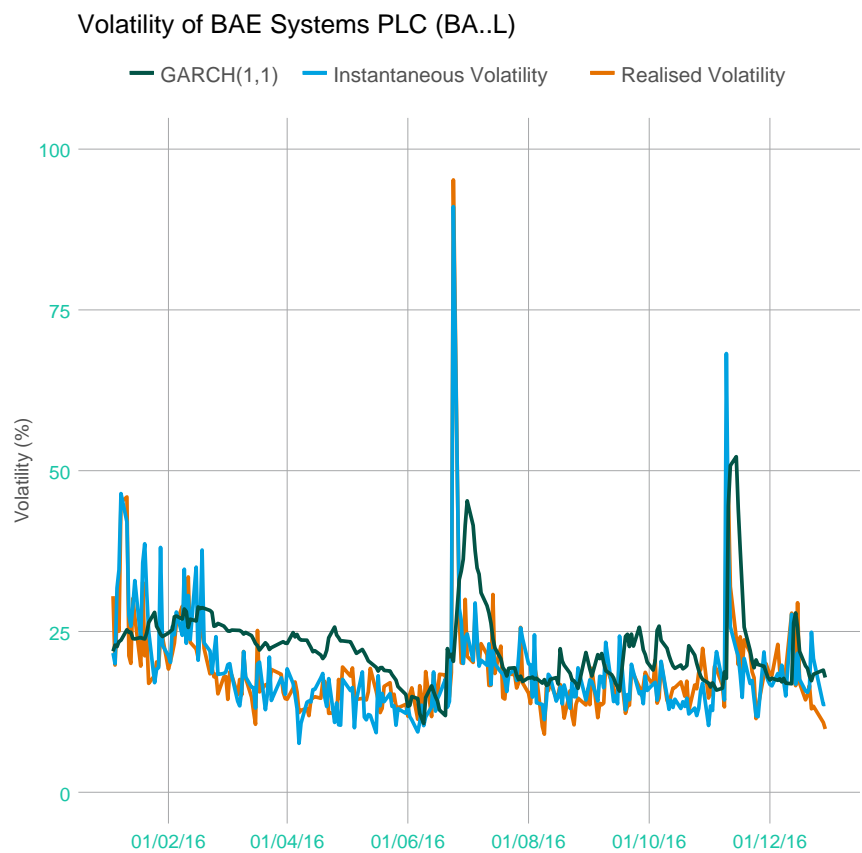}
        \caption{BA..L realised volatility (orange line) compared to one-day-ahead GARCH(1,1)
        estimation (green line) and 5 min instantaneous forecast (blue line).
        Volatilities are expressed in annualised terms.}
    \label{figure:instantaneous}
\end{figure}

Fig.\ref{figure:instantaneous} compares instantaneous volatility
to realised volatility and one-day-ahead forecast of GARCH(1,1)
model package for R by RMetrix to perform these calculations
(\cite{fgarch2013}). A sharp peak in the middle of the volatility
chart for BAE Systems corresponds to the shock on the markets
after the Brexit EU referendum results were announced on 24th
June. Instantaneous volatility correctly reflected the volatility
increase during this day. GARCH results are lugging sharp peaks of
this kind, resulting in a volatility peak the next day.

Quantitatively the volatilities could be compared using the mean
square error, defined by formula
\begin{equation}
    MSE_I=\frac{1}{N}  \sum_i \left(\sigma_{i,R} - \sigma_{i,I} \right)^2,
\label{mse}
\end{equation}
where the summation is performed over N trading days of the year,
$\sigma_{i,R}$ is the realised volatility on day $i$ and
$\sigma_{i,I}$ is the instantaneous volatility on the same day. In
a similar fashion, mean square error for GARCH(1,1) prediction
could be calculated. For a fair comparison of data shown on Fig.4,
the outlier at 24th June was removed. Overall, for 2016,
$MSE_{I}=7.1\times 10^{-6}$ although for the GARCH(1,1) model
$MSE_{GARCH}=1.7\times 10^{-3}$, more than two hundred times
larger, making its estimation less reliable. The difference is
visible on the chart: GARCH overestimated volatility in the first
part of the year and then every time the volatility had a spike,
GARCH would have similar splash next day. The instantaneous
volatility does not have this lagging factor because it uses the
same day data.

\begin{figure}[htp]
    \centering
    \includegraphics[width=0.7\textwidth, height=0.65\textwidth]{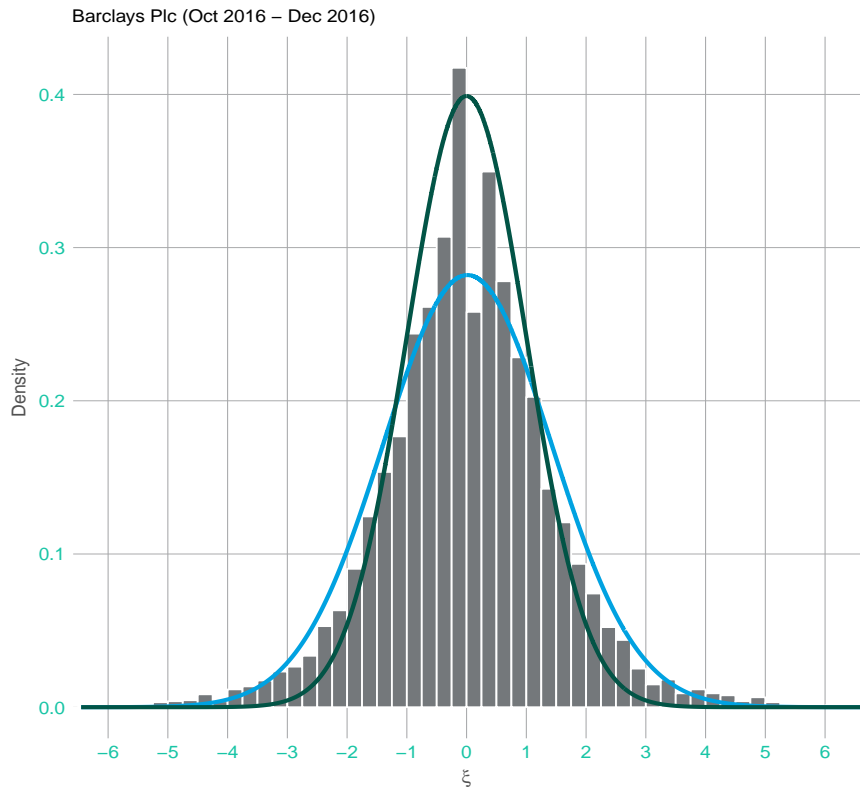}
        \caption{The distribution of random variable $\xi_i$ when historical
        5 min data was used to predict standard deviation of next 5 min price return.
        Blue line represents the fitted normal distribution
        with $\sigma(\xi)=1.454$, green line is $N(0, 1)$ distribution.}
    \label{figure:xi_distr}
\end{figure}

The ultimate test for the formula (\ref{volatility}) could be a
direct comparison of predicted volatility value with realised
volatility. Assuming that the volatility will not change in a
short time interval, one can use $ \sigma_{i,I}$ as a volatility
prediction for a future time interval. Let us introduce a random
variable

\begin {equation}
\xi_i=  \frac{r_i}{\sigma_{i-1,I}},
\label{norm_random}
\end {equation}
where an observed log return of the price is divided by the
instantaneous volatility calculated on the previous time step.
Since we divided the price return by its projected standard
deviation, the distribution of random variable $\xi$ should be
equal or comparable to normal distribution $N(0,1)$. Practical
calculation of such distribution for a liquid stock when 5 minute
price move is predicted by 5 minute observation is shown on
Fig.\ref{figure:xi_distr}. The standard deviation of normalised
returns $\sigma_{\xi}=1.454$ means that instantaneous volatility
underestimated the realised volatility by approximately 45\%.

There are two aspects  to this: first of all, the theoretical
value $\sigma_{\xi}=1$  describes well the distribution of small
returns; whereas realised $\sigma_{\xi}$ has wider tails and will
work better for larger price moves. Secondly, the distribution of
normalised returns (\ref{norm_random}) is clearly non-Gaussian
because the microstructure bias takes place on these time
intervals. It is clearly seen from volatility signature plots used
in \cite{andresen2000}: the realised volatility calculated on
short timeframes could significantly overestimate the volatility
measure even for liquid instruments.

\begin {table} [ht]

\caption {The realised standard deviation $\sigma_{\xi}$
calculated on different historical intervals (rows) used for
different forecast periods (columns).}

\centering

\begin{tabular} {| l | c | c | c | c | c |}

\hline \hline

History / Forecast          & 1 min & 5min  & 10 min           & 30 min           & 60 min \\

\hline

1 min    & 3.045 & 2.734 & 3.539 &         2.316    & 2.777 \\

5 min    &                & 1.454       & 1.446 & 1.434 & 1.428 \\

10 min &      &       &      1.363    & 1.364 & 1.361 \\

30 min &      &       &       &          1.343    & 1.279 \\

60 min  &     &       &       &       & 1.380 \\

\hline \hline

\end {tabular}

\label{table::history_forecast}

\end{table}

Around 10000 one-minute observations for Barclays Plc stock in
October 2016 were collected and combined  to construct Table
\ref{table::history_forecast}. This table demonstrates that the
short-term volatility prediction works, but it's accuracy is
limited. For this particular instrument a 5-10 min observation is
enough to estimate the volatility of a short-term price move
although one-minute historical data is not enough to make a
reliable volatility prediction.

%%\section{Liquidity of ETF}
%\input{section_7.tex}

%\section{Conclusions}
\section{Conclusions}
\label{conclusions}

We have provided a new way of short-term volatility estimation. It
is based on a market invariant which was discovered empirically
during work on algo models. The invariant represents a fundamental
property of the market and links the volatility of the instrument
with traded volume, spread size and the volume in the order book.
It was tested for a variety of stocks from different countries,
fungible instruments traded in the UK, derivatives. It is shown
that the invariant holds for liquid instruments. The market
invariant works in the state of a market equilibrium; if the
traded instrument is under a stressed condition, the deviation
from the obtained formula could be observed. Potentially, the
invariant could be distorted by unusual exchange rules or
practices, but we did not observe such markets. Another potential
correction which we could think of is a correction related to
hidden liquidity which could be easily incorporated into
equations.

The formula for instantaneous volatility is derived from the
invariant. Using realised volatility it was compared to GARCH(1,1)
estimation. The comparison showed that instant volatility is
accurate in estimating short time price volatility and correctly
predicts anomalies on market, which could arise from announcements
and geopolitical events.

The instantaneous volatility could be used for an accurate
volatility prediction in algo trading and for VIX traders. The
invariant could be also used as an indicator for instruments in a
distress condition.

\appendix

%%\section{Widely used volume based liquidity measures}
%\input{section_a.tex}

%%\section {Liquidity of a basket, test cases}
%\input{section_b.tex}

%% References
%%

\bibliographystyle{elsarticle-num}

\end{document}